\documentclass{article}
\usepackage[cmex10]{amsmath}
\usepackage{cite}
\usepackage{mathtools}
\DeclarePairedDelimiter{\ceil}{\lceil}{\rceil}
\DeclarePairedDelimiter{\floor}{\lfloor}{\rfloor}
\usepackage[ruled,linesnumbered]{algorithm2e} 
\usepackage{mdwmath}
\usepackage{mdwtab}

\usepackage{graphicx}
\usepackage{lipsum}
\usepackage{multicol}
\usepackage{multirow}
\usepackage{makecell}
\usepackage{mathtools}
\usepackage{subcaption}
\usepackage{rotating, graphicx}
\usepackage[pdftex]{color}

\usepackage[bottom]{footmisc}
\usepackage{balance}

\begin{document}

\title{Dynamic Loop Scheduling Using \\MPI Passive-Target Remote Memory Access}

\author{Ahmed Eleliemy and Florina M. Ciorba\\
	Department of Mathematics and Computer Science\\
	University of Basel, Switzerland}

\maketitle
\clearpage
\tableofcontents
\clearpage

 % !TEX root =  full_paper.tex
\begin{abstract}
\label{sec:abstract}
Scientific applications often contain large computationally-intensive parallel loops. 
Loop scheduling techniques aim to achieve load balanced executions of such applications. 
For \mbox{distributed-memory} systems, existing dynamic loop scheduling~(DLS) libraries are typically \mbox{MPI-based}, and employ a \mbox{master-worker} execution model to assign \mbox{variably-sized} chunks of loop iterations.
The \mbox{master-worker} execution model may adversely impact performance due to the \mbox{master-level} contention. 
This work proposes a \emph{distributed \mbox{chunk-calculation}} approach that does not require the \mbox{master-worker} execution scheme. 
Moreover, it considers the novel features in the latest MPI standards, such as \emph{\mbox{passive-target} remote memory access}, \emph{\mbox{shared-memory} window} creation, and \emph{atomic read-modify-write} operations.
To evaluate the proposed approach, five \mbox{well-known} DLS techniques, two applications, and two heterogeneous hardware setups have been considered. 
The DLS techniques implemented using the proposed approach outperformed their counterparts implemented using the traditional \mbox{master-worker} execution model.       


\end{abstract}

\clearpage

% !TEX root =  full_paper.tex
\section{Introduction}
\label{sec:intro} 

Over the past decade, the increasing demand for computational power of scientific applications played a significant role in the development of modern high performance computing~(HPC) systems.
The advancements of modern HPC systems at both, hardware and software levels, raise questions regarding the benefits of these advantages for successful algorithms and techniques proposed in the past.
Algorithms and techniques may, therefore, need to be revisited and \mbox{re-evaluated} to fully leverage the capabilities of modern HPC systems.
Dynamic loop scheduling~(DLS) techniques are important examples of successful scheduling techniques proposed over the years.
The DLS techniques are critical for scheduling parallel loops that are the main source of parallelism in scientific applications~\cite{fang1990dynamic}.
A large body of work on DLS was introduced between the late 1980's and the early \mbox{2000's~\cite{SS,GSS,TSS,FAC,WF,AWF,PLS}}.
These DLS techniques were used in several scientific applications to achieve a load balanced execution of loop iterations.
%\ahmedC{Execution of application's loop iterations needs to be balanced across all processing elements to achieve the best performance.}
Several factors can hinder such a load balanced execution, and consequently, degrade applications' performance. 
Specifically, problem characteristics, \mbox{non-uniform} input data sets, as well as algorithmic and systemic variations %lead to different execution times of the individual loop iteration. % which may cause a performance degradation.
lead to different execution times of each loop iteration. 
The DLS techniques are designed to mitigate load imbalance due to the aforementioned factors.

Dynamic loop \mbox{self-scheduling-based} techniques such as, \mbox{self-scheduling}~(SS)\cite{SS}, guided \mbox{self-scheduling}~(GSS)~\cite{GSS}, trapezoid \mbox{self-scheduling}~(TSS)~\cite{TSS}, factoring~(FAC)~\cite{FAC}, and weighted factoring~(WF)~\cite{WF}, constitute an important category of DLS techniques.
The distinguishing aspect of loop \mbox{self-scheduling} is that whenever a processing element becomes \emph{available} and \emph{requests work}, it obtains a collection of loop iterations (called a chunk) from a \emph{central work queue}.
Each DLS technique uses a certain function to calculate chunk sizes. 
Several implementations of \mbox{self-scheduling-based} techniques~\cite{DHTSS,DLBLtool,LBtool,HLS,PLS} employ the \mbox{master-worker} execution model, and use the classical \mbox{two-sided} MPI communication model.
In the \mbox{master-worker} execution model, a processing element~(called master) holds all the information required to calculate the chunks and serves work requests from other processing elements~(called workers).
Workers request new chunks once they become available, according to the \mbox{self-scheduling} principle.
The \mbox{master-worker} execution model highlights an important performance-relevant detail concerning the implementation of \mbox{self-scheduling-based} techniques;
it centralizes the \mbox{chunk-calculations} at the master. 
This centralization renders the master process a performance bottleneck in three scenarios:
(1)~the master process has a decreased processing capabilities; this may happen in heterogeneous systems,
(2)~the master process receives a large number of concurrent work requests; this may happen in large-scale \mbox{distributed-memory} systems,
and (3)~a combination of~(1) and~(2) may be the case when executing on \mbox{large-scale} heterogeneous systems.       
%The \mbox{self-scheduling} techniques implemented on \mbox{shared-memory} systems can naturally distribute the \mbox{chunk-calculation} among the processing elements. 
%The centralization of the \mbox{chunk-calculation} raises an important question regarding the effect of the processing capabilities of the chosen master processor on the entire performance of the \mbox{self-scheduling-based} techniques, especially on heterogeneous computing resources.

The current work proposes an approach to address the first execution scenario described above. 
Intuitively, this scenario can be avoided by mapping the master process to the processing element with the highest processing capabilities.
However, such a mapping may not always be guaranteed or feasible. 
For instance, system variations during applications' execution may adversely affect the computation or the communication capabilities of the master leading to performance degradation.  
%Vital: nobody lives or dies if self-scheduling calculates centrally or distributed.
%Moreover, the scalability of the \mbox{master-worker} execution model is known to be limited due to contention on the master with concurrent requests from a large number of workers.

%, development approach.
%Similar to \ahmedC{PGAS-based} approaches, 
%MPI introduced remote memory access~(RMA), also referred to as \mbox{one-sided} communication, in the \mbox{MPI-2} standard~\cite{MPIForum}. 
The \mbox{MPI-2} standard~\cite{MPIForum} introduced remote memory access~(RMA) semantics that were seldomly adopted in scientific applications because they had several issues~\mbox{\cite{bonachea2004problems}}.
 For instance, the unrestricted allocation of \mbox{exposed-memory} regions makes the efficient implementation of \mbox{one-sided} functions extremely difficult. 
The MPI RMA model was significantly revised in the \mbox{MPI-3} standard and more \mbox{performance-capable} RMA semantics were introduced~\mbox{\cite{RMAhoefler,zhao2016scalability}}. 
For instance, \texttt{MPI\_Win\_Allocate} was introduced to restrict the memory allocation and to allow more efficient MPI \mbox{one-sided} implementations.
The performance assessment of MPI RMA-based approaches has recently gained increased attention for different scientific applications~\mbox{\cite{zhou2016asynchronous,hammond2014implementing,shan2017experiences}}.

The present work proposes a novel approach for developing DLS techniques to execute scientific applications on \mbox{distributed-memory} systems. % that takes advantage of modern HPC systems at hardware and software levels. 
The proposed approach distributes the \mbox{chunk-calculation} of the DLS techniques among all processing elements.
Moreover, the proposed approach is implemented using the recent features offered by the \mbox{MPI-3} standard, such as \mbox{passive-target} synchronization, \mbox{shared-memory} window creation, and atomic \mbox{read-modify-write} operations. %to distribute the chunk calculations of DLS techniques among all MPI processes.
The present work is significant for improving the performance of various scientific applications, such as \mbox{N-body} simulations~\cite{fractiling}, \mbox{Monte-Carlo} simulations, and computational fluid dynamics~\cite{AWF}, that employ DLS techniques on heterogeneous and large scale modern HPC systems.
Moreover, the present work provides insights for improving existing DLS libraries~\cite{DLBLtool,LBtool} such that they take advantage of modern HPC systems by exploiting the remote direct memory access~(RDMA) capabilities of modern interconnection networks.
%At the hardware level, it exploits the increasing capabilities of remote direct memory access~(RDMA) of modern interconnection networks, whereby the main memory can be directly and remotely accessed with low latency and high bandwidth~\mbox{\cite{infiniband,omnipath}}. 
%At the software level, it considers the recent features offered by the \mbox{MPI-3} standard, such as passive-target synchronization, \mbox{shared-memory} window creation, and the atomic \mbox{read-modify-write} operations. 
%
%The focus of the current work is on assessing the performance of the proposed approach on heterogeneous resources. 
%A study of the scalability of the proposed approach would exceed the scope of this work and is planned as future work.

The main contributions of this work are: %can be summarized as follows.
(1)~Proposal and implementation of five DLS techniques with distributed \mbox{chunk-calculation} for \mbox{distributed-memory} systems. 
%(2)~Elimination of the use of the traditional \mbox{master-worker} execution model for implementing DLS techniques.
(2)~Evaluation of the benefit of using MPI \mbox{one-sided} communication and \mbox{passive-target} synchronization mode to implement the DLS techniques:~SS~\cite{SS}, GSS~\cite{GSS}, TSS~\cite{TSS}, FAC~\cite{FAC}, and~WF~\cite{WF} against other existing approaches~\cite{DLBLtool,LBtool,Eleliemy:hpcc}.

The remainder of this work is organized as follows. 
Section~\ref{sec:background} contains a review of the selected DLS techniques, as well as of the relevant research efforts on the different implementation approaches of DLS techniques for \mbox{distributed-memory} systems. 
Also, the background on the MPI RMA model is presented in Section~\ref{sec:background}.
The proposed distributed \mbox{chunk-calculation} approach and its execution model are introduced in Section~\ref{sec:main}.
The design of experiments and the experimental results are discussed in Sections~\ref{sec:DoE} and \ref{sec:results}, respectively.
The conclusions and the potential future work are outlined in Section~\ref{sec:conc}. 
%Section~\ref{sec:repro} provides the information required for the reproduction of this work.

% !TEX root =  full_paper.tex
\clearpage
\section{Background and Related Work}
\label{sec:background}

%The first part of this section offers a detailed overview of the static and dynamic loop self-scheduling techniques, as well as details of the DLS techniques \ahmedC{considered in} this work.
%The second part outlines the evolution of the implementation of the DLS techniques, with a focus on the DLS techniques for executing applications on \ahmedC{\mbox{distributed-memory}} systems.
%The last part offers an overview of the MPI RMA model used in this study.

%\subsection{Background on scheduling}
\paragraph*{\emph{\textbf{Loop Scheduling}}}
Loops are the richest source of parallelism in scientific applications~\cite{fang1990dynamic}.
\mbox{Computationally-intensive} scientific applications spend most of their execution time in large loops.
Therefore, efficient scheduling of loop iterations benefits scientific applications' performance. 
In the current work, five \mbox{well-known} DLS techniques: \mbox{self-scheduling}~(SS)~\cite{SS}, guided \mbox{self-scheduling}~(GSS)~\cite{GSS}, trapezoid \mbox{self-scheduling}~(TSS)~\cite{TSS}, factoring~FAC~\cite{FAC}, and weighted factoring~(WF)~\cite{WF}  are considered.
These techniques are considered  due to their competitive performance in different applications, and due to the fact that they are at the basis of other DLS techniques.
For instance, trapezoid factoring \mbox{self-scheduling}~(TFSS)~\cite{DSS2} is based on TSS and FAC, while adaptive weighted factoring~(AWF)~\cite{AWF} and its variants~\cite{AWFBC} are derived from FAC.
\begin{table}[!b]
	\centering
	\caption{Notation used in the present work}
	\label{tab:sym}
	%\resizebox{\columnwidth}{!}{
	\begin{tabular}{@{}l|l@{}}
		\textbf{Symbol} & \textbf{Description}                     \\ \hline 
		$N$      	& Total number of loop iterations  \\  
		$P$      	& Total number of processing elements   \\ 
		$S$      	& Total number of scheduling steps  \\  
		$B$      	& Total number of scheduling batches \\ 
		$i$  		& Index of current scheduling step, $0 \leq i \le S-1$\\  
		$b$        	& Index of currently scheduled batch, $0 \leq b \le B-1$\\ 
		$R_{i}$   	& Remaining loop iterations after \mbox{$i$-th} scheduling step\\  
		$S_{i}$   	& \makecell[l]{Scheduled loop iterations after \mbox{$i$-th} scheduling step \\ $S_{i} + R_{i} = N$} \\  
		$lp_{\text{start}}$ &\makecell[l]{Index of currently executed loop iteration, \\ $0\leq lp_{\text{start}}\leq N-1$}  \\
		$K_{0}$	&  Size of the largest chunk \\  
		$K_{S-1}$	&  Size of the smallest chunk \\  
		$K_{i}$   	& Chunk size calculated at scheduling step $i$ \\  
		$p_{j}$ 	& Processing element  $j$, $0 \leq j \le P-1$\\
		$Wp_{j}$ 	& \makecell[l]{Relative weight of processing element $j$, $0 \leq j \le P-1$, \\$\sum_{j=0}^{P-1} Wp_{j}=P$}\\  
		%$h$ 		& Average scheduling overhead for assigning a single iteration\\  
		$\sigma$  	& Standard deviation of the loop iterations' execution times\\  
		$\mu$  	& Mean of the loop iterations' execution times\\		%we donot use mean of loop iteration execution time for FS we use h assigning overhead
		$T_{\text{p}}$ 	& Parallel execution time of the entire application  \\
		$T_{\text{p}}^{\text{loop}}$ 	&  Parallel execution time of the application's parallelized loops  
	\end{tabular}
	%} % end resize box
\end{table}
 
The common aspect of all selected DLS techniques is that new chunks of iterations are assigned to processing elements once they become \emph{available} and \emph{request work}.
However, each of these DLS techniques employs a different function to calculate the size of the chunk to be assigned. 

The notation used in the present work is given in Table~\ref{tab:sym}.
In the SS~\cite{SS} technique, the assigned chunk,~$K_i$, is always~1~loop iteration. 
Due to the \mbox{fine-grained} partitioning of the loop iterations, SS can achieve a highly load balanced execution.
However, it incurs a high scheduling overhead. 
GSS~\cite{GSS}, TSS~\cite{TSS}, and FAC~\cite{FAC} outperform SS in terms of scheduling overhead, by assigning chunks of decreasing size. 
In each scheduling step~$i$, GSS uses a \mbox{non-linear} function to calculate the chunk sizes. 
It divides the remaining loop iterations,~$R_{i}$, by the total number of processing elements,~$P$.

TSS~\cite{TSS} employs a simplistic linear function to calculate the decreasing chunk sizes using a fixed decrement.
This linearity results in low scheduling overhead in each scheduling step.

FAC~\cite{FAC} schedules the loop iterations in batches of $P$ \mbox{equally-sized} chunks. 
The initial chunk size of FAC is smaller than the initial chunk size of GSS.
If more time-consuming loop iterations exist at the beginning of the loop, GSS may not balance their execution as efficiently as FAC.
Unlike GSS and TSS, which calculate the chunks deterministically, the \mbox{chunk-calculation} in FAC evolved from comprehensive probabilistic analyses. 
To calculate an optimal chunk size, FAC requires prior knowledge about the standard deviation,~$\sigma$, of loop iterations' execution times and their mean execution time~$\mu$.
A practical implementation of FAC, denoted FAC2, does not require $\mu$ and $\sigma$,  and assigns half of the remaining work in every batch~\cite{FAC}.
FAC2 evolved from the probabilistic analysis that gave birth to FAC,  and is considered in the current~work.

WF~\cite{WF} uses the FAC function to calculate the batch size. However, the processing elements execute \mbox{variably-sized} chunks of this batch according to their relative weights. 
The processor weights,~$Wp_j$, are determined prior to applications' execution and do not change during the execution.
The \mbox{chunk-calculation} function of each technique is shown in Table~\ref{tab:fun}.

\begin{table}[!b]
	\centering
	\caption{\mbox{Chunk-calculation} per loop \mbox{self-scheduling} technique}
	\label{tab:fun}
	\begin{tabular}{l | l@{}}
	 \textbf{Technique} &  \textbf{Chunk-calculation}  \\   \hline \rule{0pt}{15pt}
		STATIC      & $K_i = \ceil*{\frac{N}{P}}$.  \\ \rule{0pt}{10pt}
		SS      & $K_i=1$.   \\  \rule{0pt}{15pt}
		GSS & $K_i = \ceil*{\frac{R_{i}}{P}}$,  $R_{0} = N$. \\ \rule{0pt}{15pt}
%old way	TSS &    \begin{tabular}[l]{@{}l@{}} $K_{i}= K_{i-1} - \floor{ \frac{K_0-K_{S -1}}{S-1} }$,\\$S = \ceil{\frac{2\cdot N}{K_0+K_{S-1}}}, K_0=\ceil{ \frac{N}{2\cdot P} }\-\-, K_{S-1}=1$ \end{tabular}\\ \rule{0pt}{22pt}
		TSS & \makecell[l]{$K_{i}= K_{i-1} - \floor*{ \frac{K_0-K_{S -1}}{S-1} }$, where \\ $S = \ceil*{\frac{2\cdot N}{K_0+K_{S-1}}}$, and \\ $K_0=\ceil*{ \frac{N}{2\cdot P} }\-\-, K_{S-1}=1$.} \\ \rule{0pt}{15pt}
%very old way	FAC2 & \begin{tabular}[l]{@{}l@{}} $K_i= \ceil{ \frac{R_{i}}{2 \cdot P}}$,  $\forall \ i \mod P = 0$\\    $R_{i-1}$, otherwise \end{tabular} \\ \rule{0pt}{20pt}
% old		FAC2 &  \makecell[l]{$K_i= \ceil{ \frac{R_{i}}{2 \cdot P}}$,  $\text{if} \ i \mod P = 0$\\    $K_i=R_{i-1}$, otherwise}  \\ \rule{0pt}{15pt}
	 	FAC2 &  \makecell[l]{$K_i = \left\{ \begin{array}{ll} \ceil*{ \frac{R_{i}}{2 \cdot P}}, & \text{if} \ i\-\-\mod P = 0 \\ R_{i-1}, & \text{otherwise.} \end{array} \right.$ } \\ \rule{0pt}{15pt}
		WF   &  $K_i=\ceil*{Wp_{j}  \times K_i^{\text{FAC2}}}$.
	\end{tabular}
\end{table} 
%\min\limits_{0 \leq x \le 4} t_x
%\subsection{Related Work} 

\begin{sidewaysfigure}
	\captionsetup[subfigure]{justification=centering}
	\begin{subfigure}{0.4\textwidth}
		\centering
		\includegraphics[clip,trim=0cm 0cm 0cm 0cm, width=0.6\textwidth]{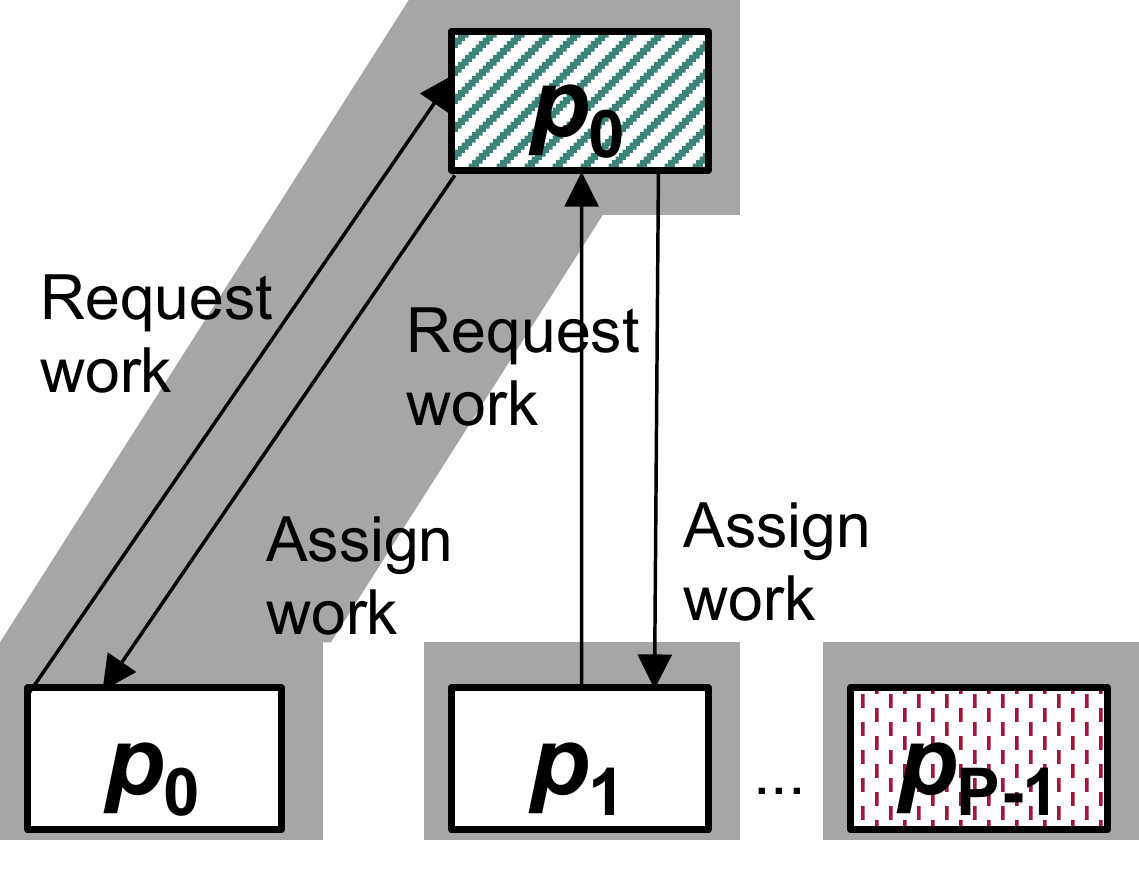}
		\subcaption{Conventional master-worker execution model}
		\label{fig:con-master-worker}%
	\end{subfigure}%
	\begin{subfigure}{0.4\textwidth}
		\centering
		\includegraphics[width=0.91\textwidth]{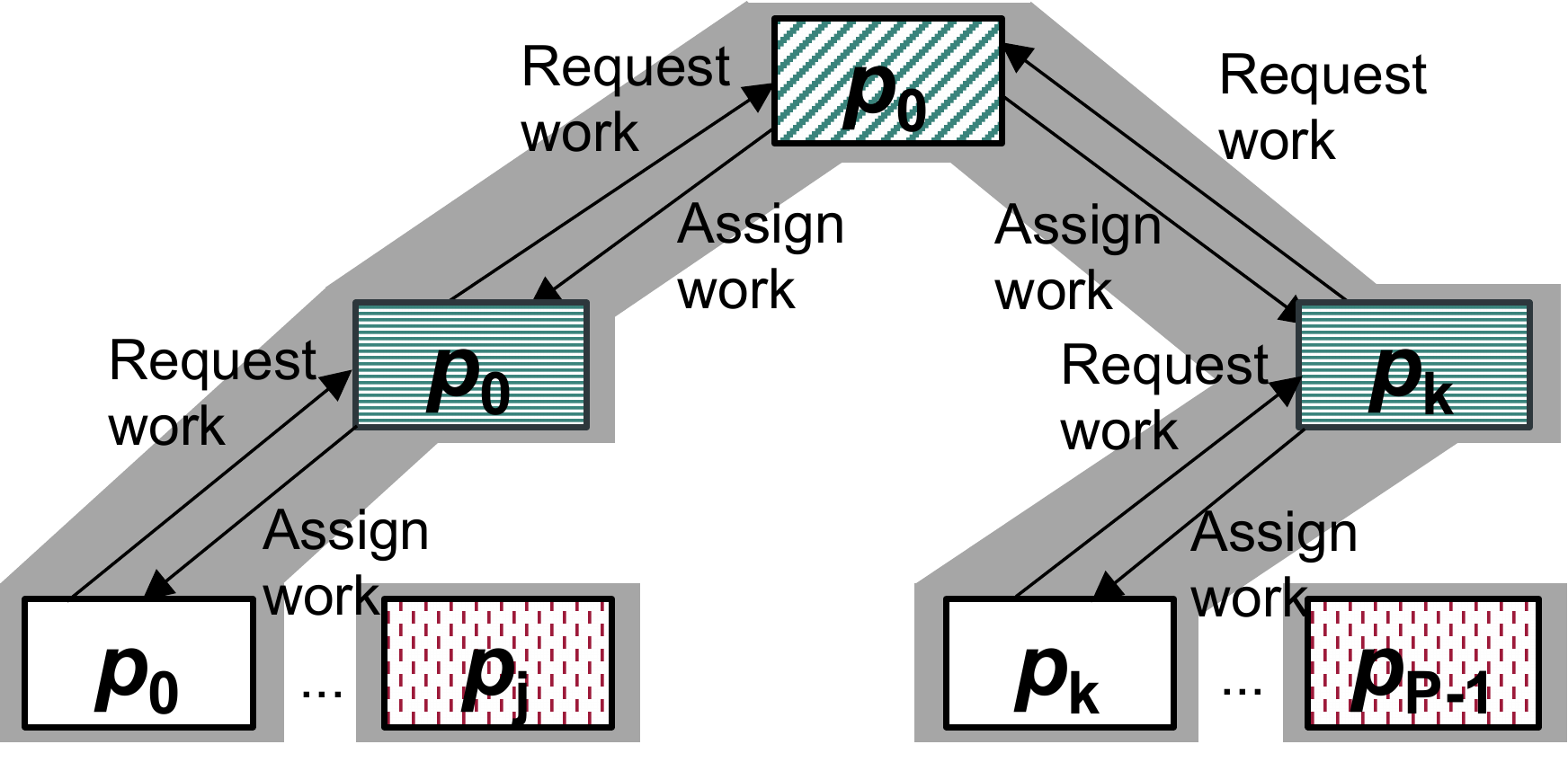}
		\subcaption{Global and local masters are located on a single physical compute node}
		\label{fig:con-h-master-same}%
	\end{subfigure}%
	\begin{subfigure}{0.4\textwidth}
		\centering
		\includegraphics[width=0.9\textwidth]{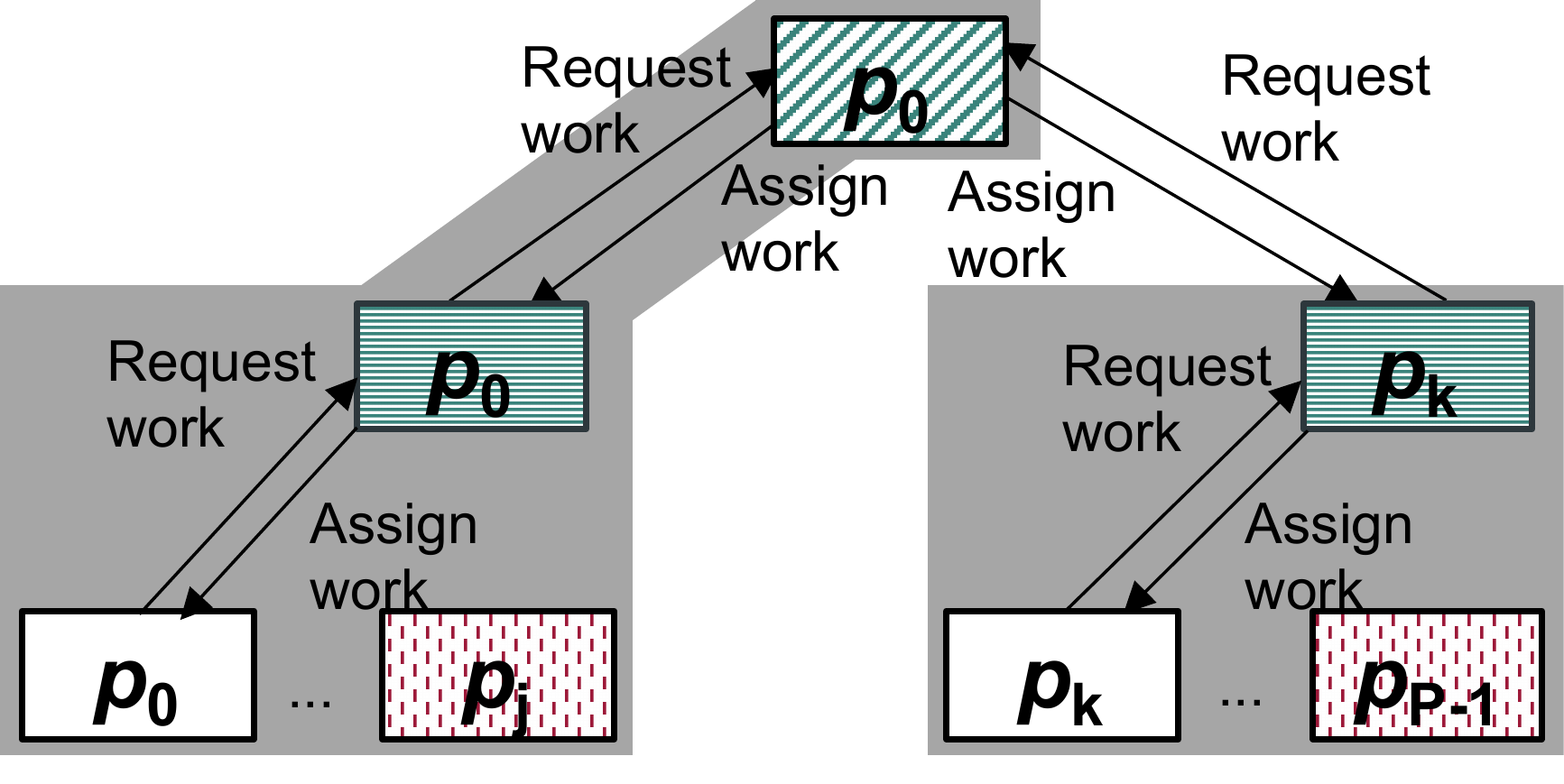}
		\subcaption{Local masters are distributed across multiple physical compute nodes}
		\label{fig:con-h-master-dis}%
	\end{subfigure}
	
	\begin{subfigure}{\textwidth}
		\centering
		\includegraphics[clip, trim=0cm 0cm 0cm 0cm,width=\textwidth]{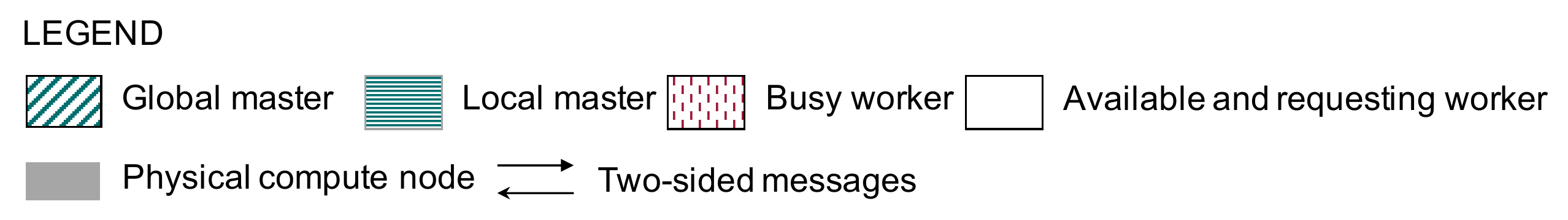}
	\end{subfigure}%
	
	\caption{Variants of the \mbox{master-worker} execution model as reported in the literature. Replication of certain processing elements is just to indicate their double role where the master participates in the computation as a worker.}
	\label{fig:models}
\end{sidewaysfigure}

\paragraph*{\emph{\textbf{Related Work}}} 
Chronopoulos et al. introduced a distributed approach for implementing self-scheduling techniques~(DSS)~\cite{DSS}.
The goal was to improve the performance of the \mbox{self-scheduling} techniques on heterogeneous and \mbox{distributed-memory} resources.
The proposed scheme was based on the \mbox{master-worker} execution model,  similar to the one illustrated in Figure~\ref{fig:con-master-worker}, 
and was implemented using the classical two-sided MPI communication.
The main idea was to enable the master to consider the speed of the processing elements and their loads when assigning new chunks.
Chronopoulos et al. later enhanced the performance of the DSS scheme using a hierarchical \mbox{master-worker} model, and proposed the hierarchical distributed \mbox{self-scheduling}~(HDSS)~\cite{DSS2} that was similar to the one illustrated in Figure~\ref{fig:con-h-master-same}.
DSS and HDSS assume a \mbox{dedicated master} configuration in which the master processing element is reserved for handling the worker requests.
Such a configuration may enhance the scalability of the proposed self-scheduling schemes.
However, it results in low CPU utilization of the master.
HDSS~\cite{DSS2} suggested deploying the \mbox{global-master} and the \mbox{local-master} on one physical computing node that has multiple processing elements to overcome the low CPU utilization of the master~(cf. Figure~\ref{fig:con-h-master-same}). 

Cari{\~n}o et al. proposed a load balancing~(LB) tool that integrated several DLS techniques~\cite{LBtool}. 
At the conceptual level, the LB tool is based on a single-level \mbox{master-worker} execution model~(cf. Figure~\ref{fig:con-master-worker}).   
However, it did not assume a \mbox{dedicated-master}. 
It introduced the \textit{breakAfter} parameter which is \mbox{user-defined} and indicates how many iterations the master should execute before serving pending worker requests.
This parameter is required for dividing the time of the master  between computation and servicing of worker requests. 
The optimal value of this parameter is application- and \mbox{system-dependent}.
The LB tool also employed the classical \mbox{two-sided} MPI communication. 

Banicescu et al. proposed a dynamic load balancing library~(DLBL) for cluster computing~\cite{DLBLtool}.
The DLBL is based on a parallel runtime environment for multicomputer applications~(PREMA)~\cite{PREMA}. 
Even though the DLBL was the first library to utilize MPI one-sided communication, the active message synchronization offered by PREMA required a \mbox{master-worker} model.
In DLBL, the master expects work requests. 
Then, it calculates the size of the chunk to be assigned and, subsequently, calls a handler function on the worker side.
The worker is responsible for obtaining the data of the new chunk from the master without any further involvement from the master.

Hierarchical loop scheduling~(HLS)~\cite{HLS} was one of the earliest efforts to utilize a hybrid MPI and OpenMP programming model to implement DLS techniques.  
HLS employed a hierarchical \mbox{master-worker} execution model, and was implemented using the classical \mbox{two-sided} MPI communication and OpenMP threads. 
Unlike HDSS~\cite{DSS2}, HLS distributes the local masters across all physical computing nodes~(cf. Figure~\ref{fig:con-h-master-dis}). 
The local masters communicate with the global master to request a new chunk when they have no more iterations to distribute between their workers.
The main limitation of HLS is that the OpenMP worker threads distribute the work using the \texttt{\#pragma omp parallel} region without the explicit use of the \texttt{nowait} clause. 
This incurs implicit local synchronization at the OpenMP level on local masters.

\paragraph*{\emph{\textbf{MPI RMA Model}}}
In MPI, the memory of each process is  by default private, and other processes cannot directly access it.
The MPI RMA model allows MPI processes to publicly expose different regions of their memory, called \textit{windows}. 
One MPI process (origin) can directly access a memory window without any involvement of the other (target) process that owns the window.
The MPI RMA has two synchronization modes: passive- and active-target. 
In the \mbox{active-target} synchronization, the target process determines the time boundaries, called \textit{epochs}, when its window can be accessed. 
In the \mbox{passive-target} synchronization, the target process has no time limits when its window can be accessed.
The present work focuses on the \emph{\mbox{passive-target} synchronization} because it requires a minimal amount of synchronization, and it efficiently allows the greatest overlap of computation and communication.
Moreover, it yields the development of DLS techniques for \mbox{distributed-memory} systems to be very similar to their original implementations for \mbox{shared-memory} systems.

% !TEX root =  full_paper.tex
%\section{Distributed Scheduling Model}% of DLS on Distributed Memory Systems }
\section{The Proposed Approach}% of DLS on Distributed Memory Systems }
\label{sec:main}
%The \mbox{master-worker} model has a \mbox{well-known} issue concerning its scalability.
%When a master has a significant number of workers, it may have a large queue of waiting requests. 
%Consequently, new requests of workers may be delayed until the master serves all requests in the queue.
%When the master is supposed to participate in executing part of the application's workload, the issue become more challenging because the master has to spend the majority of its time handling the workers' requests. 

%This section \ahmedC{consists} of two parts: the first part describes \ahmedC{the implementation of the} chunk calculation and the chunk assignment in DLS techniques. 
%The second part introduces the distributed \ahmedC{\mbox{chunk-calculation}} approach, discusses certain aspects that \ahmedC{may limit} the proposed approach. 
%\paragraph*{\emph{\textbf{DLS Techniques Innerworkings}}} 

The main challenge to design a distributed \mbox{chunk-calculation} approach is associated with the \mbox{chunk-calculation} functions of the DLS techniques. 
To calculate the current chunk to be assigned, these functions (except for SS) require either the value of the remaining loop iterations $R_i$ or the value of the previous chunk $K_{i-1}$ (cf. Table~\ref{tab:fun}).
Therefore, the chunks have to be calculated in a sequential manner, i.e., two chunks cannot be calculated simultaneously because the values of $R_i$ and $K_{i-1}$ change after each \mbox{chunk-calculation}.  
This serialization  perfectly fits any \mbox{master-worker-based} execution approach because the master serves one request at a time, and consequently, the \mbox{chunk-calculation} can be performed in a centralized and sequential fashion.   
%For instance in GSS, the chunk $K_i$ requires the value of $R_i$ while the chunk $K_{i+1}$ requires the value $R_{i+1}$ which equals to $R_i$ - $K_i$}.
%Such sequential and centralized manner forms \emph{a challenge} for any effort to design a distributed chunk-calculation approach.
%To face this challenge, we divide the DLS techniques into two main parts: \textit{\mbox{chunk-calculation}} and \textit{\mbox{chunk-assignment}}.
%While \mbox{\textit{chunk-calculation}} requires certain computations, the \mbox{\textit{chunk-assignment}} requires manipulation of certain \textit{scheduling data} such as, $R_{i}$~and~$lp_\text{start}$.

The approach proposed in this work introduces certain transformations of the respective \mbox{chunk-calculation} functions from Table~\ref{tab:fun}, such that the \mbox{chunk-calculation} depends only on the index of the last scheduling step~$i$. 
These transformations are shown below in Equations~\ref{eq1}-\ref{eq3}.
\begin{align}
%\begin{equation}
\label{eq1}
\text{GSS:} & \ \ \ K^\prime_{i} = \left \lceil \left(\frac{P-1}{P}\right)^i \cdot \frac{N}{P} \right \rceil \\
%\end{equation}  
%\begin{equation}
\label{eq2}
\text{TSS:} & \ \ \ K^\prime_{i} = K_0 - i \cdot \left \lfloor \frac{K_0-K_{S-1}}{S-1} \right \rfloor \\
%\end{equation}
%\begin{equation}
\label{eq3}
\text{FAC2:} & \ \ \ K^\prime_{i} = \left \lceil \left(\frac{1}{2}\right)^{i_{\text{new}}} \cdot \frac{N}{P} \right \rceil, \ \ i_{\text{new}}= \left \lfloor \frac{i}{P} \right \rfloor+1
%\end{equation}
\end{align}

For GSS and FAC, the transformations were already introduced in the literature~\cite{FAC}. 	
For TSS, the mathematical derivation of the transformation is as follows.
Given that $S$, $K_0$, and $K_{S-1}$ are constants, the TSS equation in Table~\ref{tab:fun} can be represented as follows. 	
\begin{align}
\label{eq6}
&K_{i}= K_{i-1} - C, \text{where C is a constant value.}\\
\label{eq7}
& C=\floor{ \frac{K_0-K_{S -1}}{S-1} }
\end{align}	
\text{Calculating} $K_1$, $K_2$, $K_3$,  $...$ $K_i$ \text{using Equation}~\ref{eq6}\\	
\begin{align}
& K_{1} = K_{0} - C \\
&K_{2}= K_{1} - C = (K_{0} - C) - C =  K_{0} - 2 \cdot C  \\
&K_{3}= K_{2} - C = (K_{0} - 2 \cdot C) - C = K_{0} - 3 \cdot C \\
&K_{i} = K_{0} - i \cdot C\\
& K_{i} = K_{0} - i \cdot \floor{ \frac{K_0-K_{S -1}}{S-1} } = K^\prime_{i}
\end{align}
%can be found in the companion research report~\cite{myequation}.
WF uses the \mbox{chunk-calculation} function of FAC and can naturally inherit the transformed FAC function.

%For instance, \mbox{chunk-calculation} in GSS performs the following computation~$\ceil{\frac{R_{i}}{P}},$ and the \mbox{chunk-assignment} calculates the start loop index~$lp_\text{start}$ and updates the remaining loop iterations~$R_{i}$.
%Employing any variant of the \mbox{master-worker}~(cf. Section~\ref{sec:background}) model results in centralizing the execution of both, \mbox{chunk-calculation} and \mbox{chunk-assignment} at the \mbox{master-side}.
  
%When executing applications on heterogeneous resources, \ahmedC{the processor, where the master process is located must be carefully chosen}. 
%It must be the most powerful processing element in terms of computation and communication capabilities. 
%Otherwise, the limited capabilities of the master may have a strong adverse impact on the performance of \ahmedC{the} workers. 

%\paragraph*{\emph{\textbf{Proposed Distributed \mbox{Chunk-Calculation}}}}
The proposed approach uses Equations~\ref{eq1}-\ref{eq3} to \emph{distribute} the \mbox{chunk-calculation} across all processing elements.
Thus, only one processing element (called \texttt{coordinator}) stores index of the last scheduling step~$i$ and the index of the last scheduled loop iteration~$lp_{\text{start}}$. 
%Certain data manipulation is required after each \mbox{chunk-calculation} by any processing element. 
%Each processing element calculates its chunk by getting a copy of $i$ and atomically increment it by one.
%Then, two processing elements can simultaneously calculate their chunks based on their own copies of~$i$.
%To execute the calculated chunk, each processing element obtain a copy of $lp_{start}$ and atomically increment its calculated chunk size.      
%In the MPI RMA model, such global synchronization can be achieved using \mbox{\texttt{MPI\_Win\_lock}(\mbox{\texttt{MPI\_LOCK\_EXCLUSIVE}})}.
%However, such global synchronization incurs high overhead.
%To avoid this overhead, the approach proposed in this work uses \mbox{\texttt{MPI\_Win\_lock}\mbox{(\texttt{MPI\_LOCK\_SHARED})}} and certain MPI atomic operations, such as \mbox{\texttt{MPI\_Get\_accumulate}}.   
%The idea is that every processing element can obtain the index of the last scheduling step, and atomically accumulate the size of its chunk to the index of the last scheduling step.
%
%The rationale is that every processing element needs to perform three steps to obtain a new chunk, as illustrated in Figure~\ref{fig:proposed}. 

\begin{figure}[!t]
	\centering
	\includegraphics[clip, trim=0cm 1.2cm 0.1cm 0cm,width=\columnwidth]{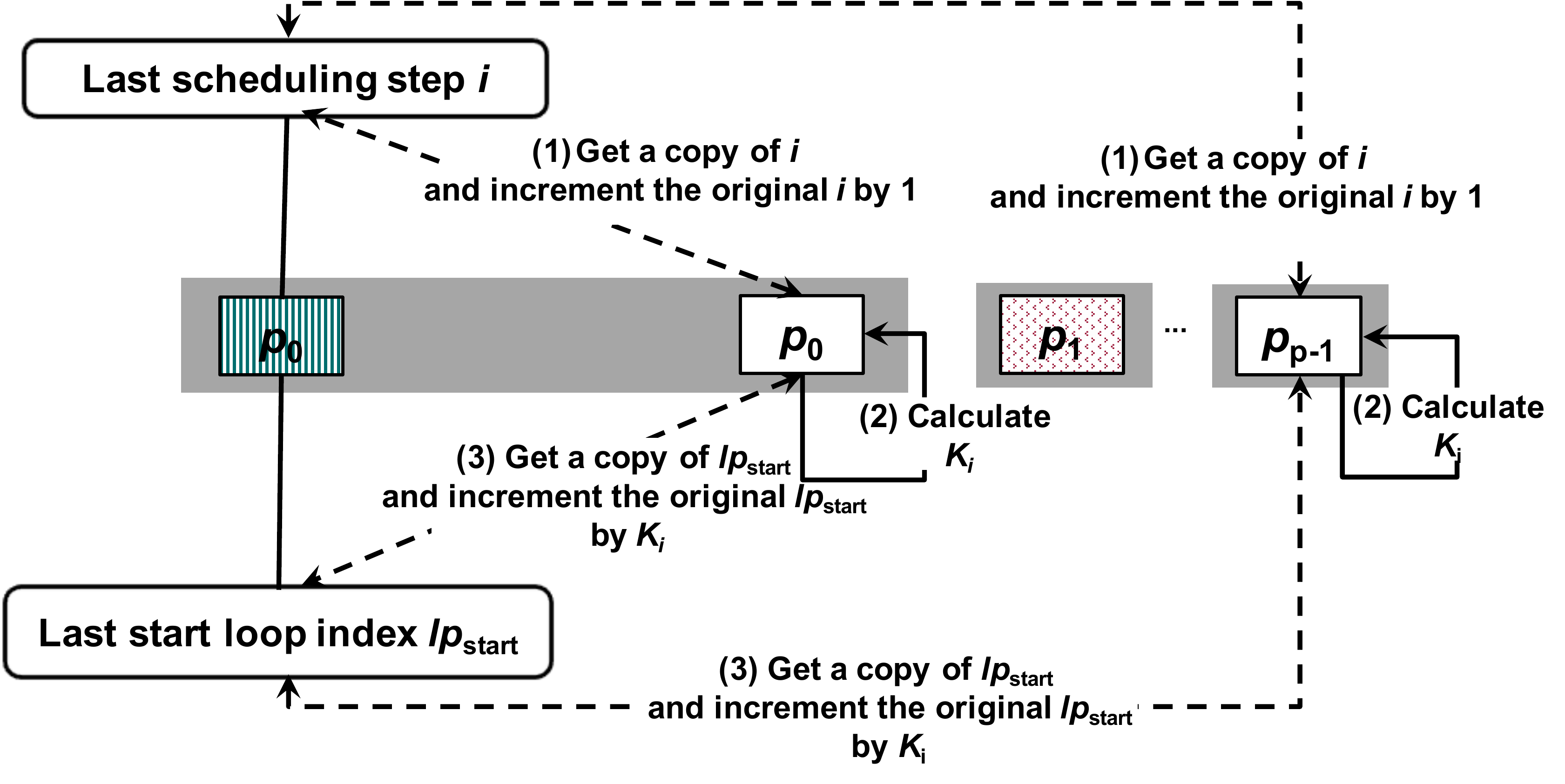}
	\includegraphics[width=\columnwidth]{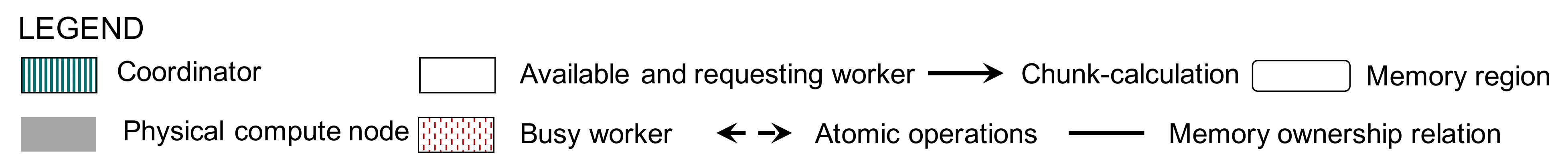}
	\caption{The proposed distributed \mbox{chunk-calculation} approach using MPI RMA and \mbox{passive-target} synchronization. }
	\label{fig:proposed}
\end{figure}

Figure~\ref{fig:proposed} illustrates the main steps of the proposed \emph{distributed \mbox{chunk-calculation}} approach:\\
\textbf{Step~1.}~the processing element~$p_j$ obtains a copy of the last scheduling step index,~$i$, and atomically increments the global $i$ by one.\\
\textbf{Step~2.}~$p_j$ \emph{only} uses its local copy of~$i$ (before the increment) to calculate $K_i$ with the function of the selected DLS technique (Equations~\ref{eq1}-\ref{eq3}).\\
\textbf{Step~3.}~$p_j$ obtains a copy of the last loop index start,~$lp_{\text{start}}$, and atomically accumulates the size of the calculated chunk,~$K_i$, into it.\\ 
Finally, $p_j$ executes loop iterations between~$lp_{\text{start}}$ (before accumulation) and~\mbox{min($lp_{\text{start}} + K_i$, $N$)}.
The atomic operations in Steps~1 and~3 guarantee the exclusive access to~$i$ and~$lp_{\text{start}}$. %which is necessary for the correctness of the chunk calculation and the chunk assignment.

The MPI RMA model provides the necessary function calls that can be used in the implementation of the proposed approach.
For instance, the coordinator MPI process can use \texttt{MPI\_Win\_create} to expose the shared variables, such as $i$ and $lp_{\text{start}}$, to all other MPI processes.
The \mbox{passive-target} synchronization mode (\texttt{MPI\_Win\_lock(MPI\_LOCK\_SHARED)}) can be used with certain MPI atomic operations, such as \mbox{\texttt{MPI\_Get\_accumulate}}, to grant the exclusive access to~$i$ and~$lp_{\text{start}}$ by all MPI processes. 
For more information regarding the implementation, the reader is referred to the code that is developed under the LGPL license and available online~\cite{mycode}.

%Given that \mbox{\texttt{MPI\_Get\_accumulate}} is an atomic operation, it is not possible to have two processing elements with the \emph{same copy of the last} scheduling step~$i$ nor the \emph{same copy of the last} start loop index~$lp_{\text{start}}$.
%The processing elements can use Equations~\ref{eq1}, \ref{eq2}, and \ref{eq3} where $i$ is the scheduling step local to \ahmedC{calculate their chunk sizes}.
%Then, each processing element uses \mbox{\texttt{MPI\_Win\_lock}} (\mbox{\texttt{MPI\_LOCK\_SHARED}}) and \mbox{\texttt{MPI\_Get\_accumulate}} to obtain a local copy of the total number of scheduled iterations $S_i$ and increases the total scheduled iterations step by the new locally calculated chunk size $C_i$. 
%Finally, each processing element executes the loop iterations with indices between $K_i$ and  $S_i + \text{min}(C_i, I-S_i )$.
%All processing elements terminate when all iterations have been executed~($lp_{\text{start}} \geq N$). 

\begin{figure}[!t]
	\centering
	\includegraphics[clip, trim=0cm 0cm 0cm 0cm ,width=\columnwidth]{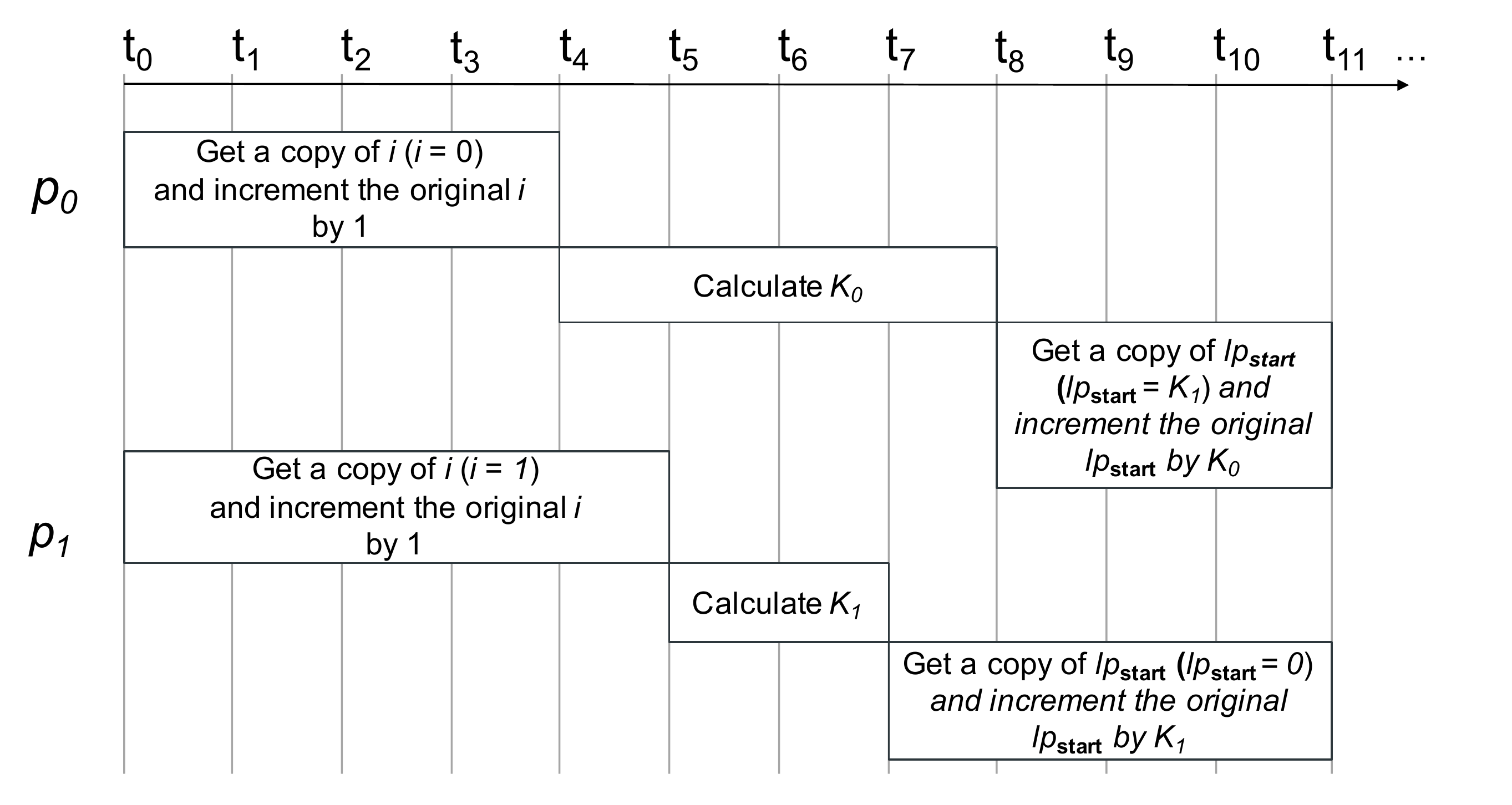}
	\caption{Schematic execution of the proposed distributed \mbox{chunk-calculation} approach on two processors that calculate one chunk each.}
	\label{fig:timing}
\end{figure}

Figure~\ref{fig:timing} illustrates the DLS execution using the proposed distributed \mbox{chunk-calculation} approach.
The calculation of chunks $K_0$ and $K_1$ is distributed between processors $p_0$ and $p_1$. 
The time required to calculate $K_0$ overlaps  with the time taken to calculate $K_1$. 
In the traditional \mbox{master-worker} execution model, there is no such overlap since all the chunk calculations are centralized and performed by the master in sequence.
The time required to serve the first work request (including \mbox{chunk-calculation} and \mbox{chunk-assignment}) delays the second work request.
Moreover, the time required to serve the work requests is proportional to the processing capabilities of the master processor, which may result in additional delays as discussed in Section~\ref{sec:intro}.

The proposed distributed \mbox{chunk-calculation} approach may result in a different ordering of assigning and executing loop iterations compared to the traditional \mbox{master-worker} execution model. 
For instance, when GSS is the chosen scheduling technique in Figure~\ref{fig:timing} and \mbox{$N=10$}, $p_0$ obtains a local copy of the last scheduling index $i=0$ at $t_4$. Also, $p_1$ obtains at $t_5$  a local copy of the last scheduling index $i=1$. 
Both, $p_0$ and $p_1$ use their copies of $i$ and calculate $K_0=5$ and $K_1=3$, respectively.
The proposed approach does not guarantee that $p_0$ and $p_1$ will execute loop iterations from $lp_{\text{start}}=0$ to $lp_{\text{start}}=4$ and $lp_{\text{start}}=5$ to $lp_{\text{start}}=7$.
Figure~\ref{fig:timing} shows the case when the \mbox{chunk-calculation} on $p_0$ is longer than on $p_1$, and results in assigning $p_1$,  loop iterations between $lp_{start}=0$ and $lp_{start}=2$, while $p_0$ is assigned loop iterations between $lp_{\text{start}}=3$ and $lp_{\text{start}}=7$.
Given that DLS techniques address, by design,  independent loop iterations with no restrictions on the monotonicity of the loop execution, the proposed approach does not affect the correctness of the loop execution.

%Algorithm~\ref{algo} shows the the proposed approach DLS techniques using the conventional \mbox{master-worker} as reported in the literature and the proposed approach.
  
%	\begin{algorithm}[h]
%	\If{Master}
%	{
		
%	}
%	\If{worker}
%	{}
%	\caption{DLS techniques using the conventional \mbox{master-worker} model}
%	\label{algo:1}
%	\end{algorithm}

%In addition to the required information to calculate the size of the new chunk  (\textit{chunk\_size}), the master has to keep track of the total scheduled loop iterations, so that it can calculate the index of the start loop iteration, and then it can do the chunk assignment. 
%This centralization may have a negative impact on the performance of the DLS technique. 

% !TEX root =  full_paper.tex
\section{Design and Setup of Experiments}
\label{sec:DoE}
In the present work, the performance of two different implementations of DLS techniques is assessed.
The first implementation, denoted~\texttt{One\_Sided\_DLS}, employs the proposed distributed \mbox{chunk-calculation} approach and uses \mbox{one-sided} MPI communication in the \mbox{passive-target} synchronization mode.
The second implementation, denoted~\texttt{Two\_Sided\_DLS}, employs a \mbox{master-worker} model and uses the \mbox{two-sided} MPI communication.
Both implementations assume \mbox{a non-dedicated} coordinator (or \mbox{a non-dedicated} master) processing element. 

%\subsection{Selected Application}
\paragraph*{\emph{\textbf{Selected Applications}}}
\label{subsec:sw_specs}
Two \mbox{computationally-intensive} parallel applications are considered in this study. 
The first application, called~PSIA~\cite{psia}, uses a parallel version of the \mbox{well-known} spin-image algorithm~(SIA)~\cite{sia}.
SIA converts a 3D object into a set of 2D images.
The generated 2D images can be used as descriptive features of the 3D object.
The second application calculates the Mandelbrot set~\cite{mandelbrot1980fractal}.
The Mandelbrot set  is used to represent geometric shapes that have the \mbox{self-similarity} property at various scales.
Studying such shapes is important and of interest in different domains, such as biology, medicine, and chemistry~\cite{mandelbrot}.

Both applications contain a single large parallel loop that dominates their execution times.
Dynamic and static distributions of the most \mbox{time-consuming} parallel loop across all processing elements may enhance applications' performance.
The pseudocodes of both applications listed in Algorithm~\ref{algo:psia} and~\ref{algo:mandel}. 

	\begin{algorithm}[!b]
		\SetKwInOut{Input}{Inputs}
		\SetKwInOut{Output}{Output}
		spinImagesKernel (W, B, S, OP, M)\;
		\Input{W: \mbox{image width}, B: \mbox{bin size}, S: \mbox{support angle}, \mbox{OP: list of 3D points}, \mbox{M: number of spin-images}}
		\Output{R: list of generated spin-images}  
		\For{ {\color{blue} i = 0 $\rightarrow$ M}}
		{ 
			P = OP[i]\;
			tempSpinImage[W, W]\;
			\For{j = 0 $\rightarrow$ $length(OP)$}
			{
				X = OP[j]\;
				$np_i$ = getNormalVector(P)\;
				$np_j$ = getNormalVector(X)\;
				\If{acos($np_i \cdot np_j$) $\le S$}
				{
					$k$  =  $\Bigg \lceil$ $\cfrac{W/2 - np_i \cdot (X-P) }{B}$ $\Bigg \rceil$\;
					\vspace{0.2cm}
					$l$ =  $\Bigg \lceil$  $\cfrac{ \sqrt{||X-P||^2 - (np_i\cdot(X-P))^2} }{B}$  $\Bigg \rceil$\;
					
					\If{0 $\le$ k $\textless$ W and 0 $\le$ l $\textless$ W}
					{ tempSpinImage[k, l]++\;	}
				}
			}
			R.append(tempSpinImage)\;
		}
		\caption{Parallel \mbox{spin-image} calculations. The main loop is highlighted in the blue color.}
		\label{algo:psia}
	\end{algorithm}

	\begin{algorithm}[!t]
		\SetKwInOut{Input}{Inputs}
		\SetKwInOut{Output}{Output}
		mandelbrotSetCalculations (W, T)\;
		\Input{W: \mbox{image width}, CT: \mbox{Conversion Threshold}}
		\Output{V: Visual representation of mandelbrot set calculations }  
		\For{ {\color{blue} counter = 0 $\rightarrow$ $W^{2}$}}
		{ 
			$x$ = counter $/$ W\;
			$y$ = counter $\mod$ W\;
			$c$= complex(x\_min + x/W*(x\_max-x\_min) , y\_min + y/W*(y\_max-y\_min))\;
			$z$ = complex(0,0) \;
			\For{$k=0 \rightarrow$ $CT$ OR $|z| \textless 2.0$}
			{
				$z = z^4 + c$\;
			} 
			
			\eIf{  $k=CT$}
			{
				set $V(x,y)$ to black\;
			}
			{
				set $V(x,y)$ to blue\;
			}
		}
		\caption{Mandelbrot set calculations. The main loop is highlighted in the blue color.}
		\label{algo:mandel}
	\end{algorithm}	

Table~\ref{tab:character} summarizes the execution parameters used for both selected applications.
These parameters were selected empirically to guarantee a reasonable average iteration execution time that is larger than 0.2 seconds.

\begin{table*}[!t]
	\centering
	\caption{Execution parameters of both selected applications}
	\label{tab:character}	
	
	\begin{tabular}{l|l|l|l}
		Application & Input Size & Output size & Other parameters~\cite{Eleliemy:hpcc, mandelbrot} \\ \hline
		\multirow{3}{*}{PSIA} & \multirow{3}{*}{800,000 3D points~\cite{dataset}} & \multirow{3}{*}{288,000  images} & 5x5 2D image \\
		&  &  & bin-size = 0.01 \\
		&  &  & support-angle = 2 \\ \hline
		\multicolumn{1}{c|}{\multirow{3}{*}{Mandelbrot}} & \multirow{3}{*}{No input data} & \multirow{3}{*}{One image} & image-width = 1152x1152 \\
		\multicolumn{1}{c|}{} &  &  & number of iterations = 1000 \\
		\multicolumn{1}{c|}{} &  &  & Z exponent = 4
		\end{tabular}
			
\end{table*}

\clearpage
\paragraph*{\emph{\textbf{Hardware Platform Specifications}}}
\label{subsec:hw_specs}
Two types of computing resources are used in this work.
The first type, denoted KNL, refers to standalone Intel Xeon Phi~7210 manycore processors with 64~cores,~\mbox{96 GB~RAM} (flat mode configuration), and 1.3~GHz CPU frequency.
The second type, denoted Xeon, refers to \mbox{two-socket} Intel Xeon~\mbox{E5-2640} processors with 20~cores,~\mbox{64 GB~RAM}, and~2.4 GHz CPU frequency.

These platform types are part of a \mbox{fully-controlled} computing cluster that consists of 26~nodes:~22 KNL nodes and 4~Xeon nodes.
All nodes are interconnected in a \mbox{non-blocking} \mbox{fat-tree} topology. 
The network characteristics are: Intel \mbox{Omni-Path} fabric, \mbox{100~GBit/s} link bandwidth, and 100~ns~network latency.
Each KNL node has \emph{one} Intel \mbox{Omni-Path} host fabric interface adapter.
Each Xeon node has \emph{two} Intel \mbox{Omni-Path} host fabric interface adapters. 
All host fabric adapters use a single PCIe~x16~100~Gbps port.
As this computing cluster is actively used for research and educational purposes, only 40\% of the cluster could be \emph{dedicated} to the present work, at the time of writing, specifically 288 cores out of the total 696 available cores.

In the present work, the total number of cores is fixed to 288 cores, whereas, the ratio between the KNL and the Xeon cores is varied. 
Two ratios have been considered:~2:1~represents the case when the KNL cores are the dominant type of computing resources, %\ahmedC{1:1} represents \ahmedC{the} case when the number of cores of Type1 and Type2 are equal,
and 1:2 represents the complementary  case where the Xeon cores are the dominant computing resources. 
Table~\ref{tab:ratios} illustrates these two ratios.
\begin{table}[!b]
	\centering
	\caption{Ratios between the KNL and Xeon cores}
	\label{tab:ratios}
	\begin{tabular}{c|c|c|c}
		Ratio & \textbf{KNL cores} & \textbf{Xeon cores} & \textbf{Total cores} \\ \hline
		2:1 &192            & 96 &288   \\ 
		1:2  & 96             & 192    &288
	\end{tabular}
\end{table}
Also, 48 KNL cores and 16 Xeon cores per node are used, while the remaining cores on each node were left for other \mbox{system-level} processes.

\paragraph*{\emph{\textbf{Mapping of the Coordinator (or the Master) Process to a Certain Core}}}
Two mapping scenarios are considered for the assessment of the proposed \mbox{\texttt{One\_Sided\_DLS}} approach vs. the \mbox{\texttt{Two\_Sided\_DLS}} approach.
In the first mapping scenario, the process that plays the role of the coordinator for \mbox{\texttt{One\_Sided\_DLS}} or the role of the master for \mbox{\texttt{Two\_Sided\_DLS}} is mapped to a KNL core.
The CPU frequency of a single KNL core is~1.3~GHz, while the CPU frequency of a single Xeon core is~2.4~GHz. 
Therefore, this mapping represents a case when the coordinator (or the master) process is mapped to one of the cores that has the lowest processing capabilities.
In the second mapping scenario, the process that plays the role of the coordinator (or the master) is mapped to a Xeon core, which is the most powerful processing element in the considered system.
Comparing the results of both scenarios shows the adverse impact of reduced processing capabilities of the master on the performance of the DLS techniques using \mbox{\texttt{Two\_Sided\_DLS}}. 
On the contrary, the same mapping for the coordinator process did not affect the performance of the DLS techniques using \mbox{\texttt{One\_Sided\_DLS}}.

%\flo{Here it is the place to mention about the two network ports for each of the 3 Xeon nodes and 1 network port for each of the 3 KNLs. Thus, this creates a heterogeneity in how many MPI processes /socket need to ``serialize'' their messages before they are injected into the non-blocking network! - If the answer is: none, this still should be included here!}
% !TEX root =  full_paper.tex
\section{Results and Discussion}
\label{sec:results}
%Figure~\ref{fig:ratio} 
%\flo{Missing intro sentence to this section...}
The straightforward parallelization~(STATIC) is used as a baseline to assess the performance of the selected DLS techniques on the target heterogeneous computing platform.
STATIC assigns $\ceil*{N/P}$ loop iterations to each processing element.
The considered implementation of STATIC follows the \mbox{self-scheduling} execution model where every worker obtains a single chunk of size~$\ceil*{N/P}$ loop iterations at the beginning of the application execution.
By employing STATIC, the percentage of the parallel execution time of the selected applications' main loops~$T_{\text{p}}^{\text{loop}}$ are 98\% and 99.4\% of the parallel execution times for PSIA and Mandelbrot, respectively.
Such high percentages show that the performance of both applications is dominated by the execution time of the main loop.
Hence, for the remaining results in this section, the analysis concentrates on the parallel loop execution time,~$T_{\text{p}}^{\text{loop}}$.
All experiments were repeated 20 times and the median results are reported in all figures.

\begin{figure*}
 	\captionsetup[subfigure]{justification=centering}
 	\begin{subfigure}{\textwidth}
 		\centering
 		\includegraphics[clip,trim=2cm 1cm 2cm 0cm,scale=0.55]{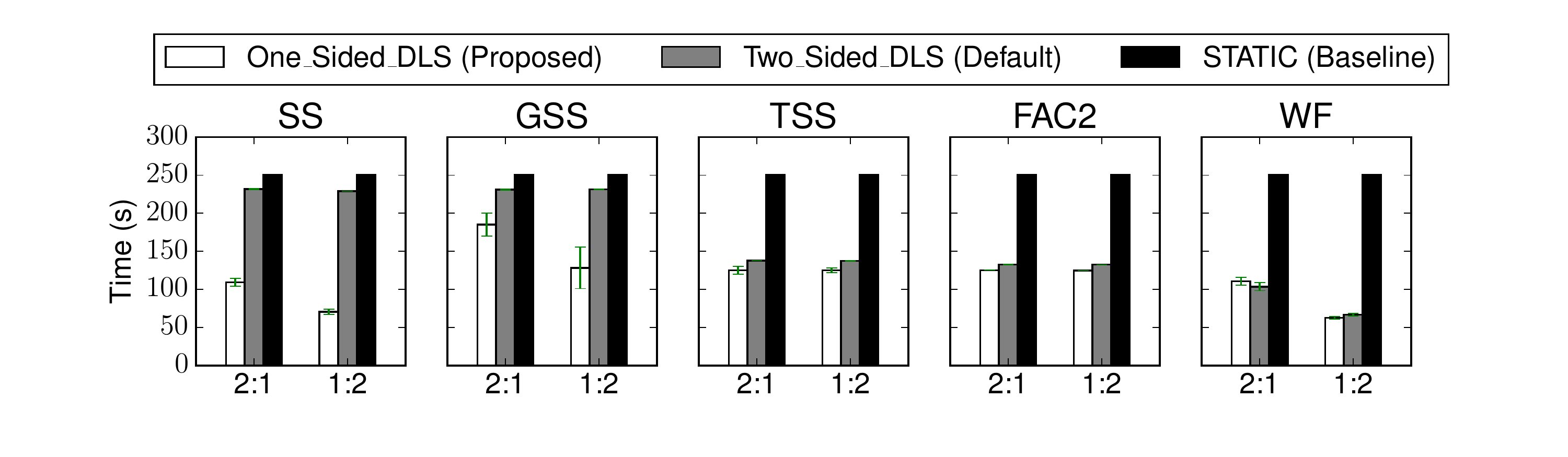}
 		\caption{The coordinator$|$master is mapped to a KNL core}
 		\label{fig:1:1}
 	\end{subfigure}
 	\\
 	\begin{subfigure}{\textwidth}
 		\centering
 		\includegraphics[clip,trim=2cm 1cm 2cm 1cm, scale=0.55]{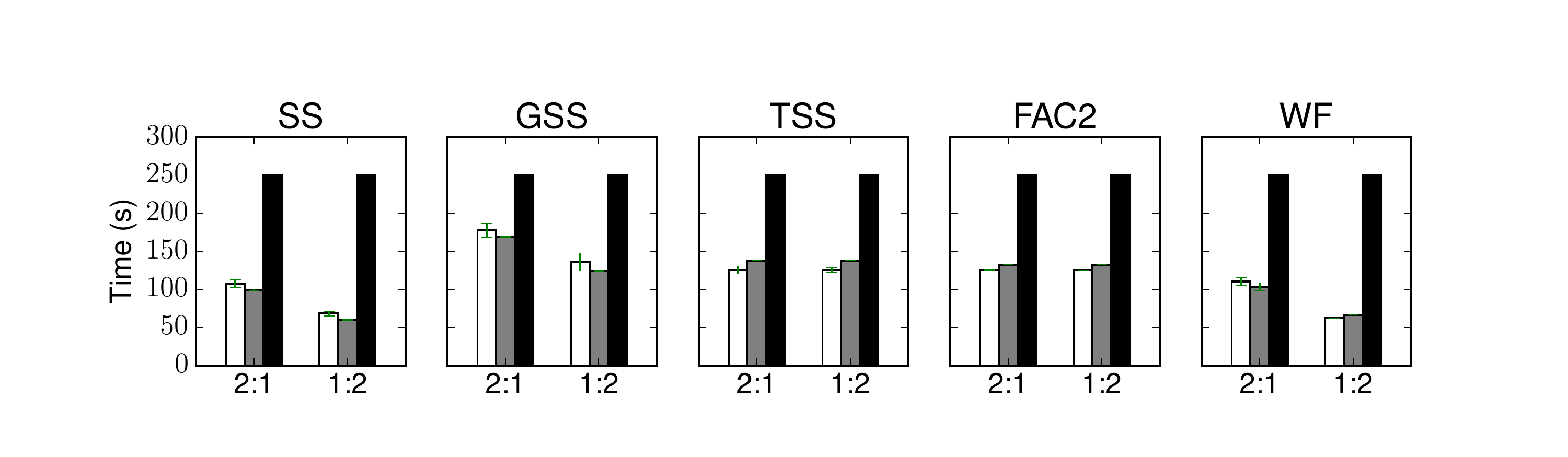}
 		\caption{The coordinator$|$master is mapped to a Xeon core}
 		\label{fig:1:2}
 	\end{subfigure}
 	\caption{Performance of the proposed approach vs. the existing \mbox{master-worker} based approach for  the PSIA. The $x$-axis represents the two ratios between the KNL cores and the Xeon cores.}
 	\label{fig:1}
 \end{figure*}

For the PSIA application, Figure~\ref{fig:1:1} shows that SS, GSS, and TSS implemented with \mbox{\texttt{One\_Sided\_DLS}} outperformed their respective versions using \mbox{\texttt{Two\_Sided\_DLS}}.
For instance, when the ratio of the KNL cores to the Xeon cores was 2:1, the parallel loop execution time,~$T_{\text{p}}^{\text{loop}}$, of SS required 109 and 233 seconds with \mbox{\texttt{One\_Sided\_DLS}} and \mbox{\texttt{Two\_Sided\_DLS}}, respectively.
Similarly, when the ratio was 2:1, the parallel loop execution time,~$T_{\text{p}}^{\text{loop}}$, of GSS and TSS increased from 185 and 125 seconds to 236 and 136 seconds, respectively.

When the ratio was 1:2, the total processing capabilities of the system increased because the number of Xeon cores increased.
However, the parallel loop execution time,~$T_{\text{p}}^{\text{loop}}$, of SS, GSS, and TSS implemented using \mbox{\texttt{Two\_Sided\_DLS}} did not take the advantage of increasing the total number of Xeon cores.
For instance using \mbox{\texttt{One\_Sided\_DLS}}, changing the ratio from 2:1 to 1:2 reduced the~$T_{\text{p}}^{\text{loop}}$ of SS from 109 to 68.5 seconds.
FAC and WF behaved similarly using both, \mbox{\texttt{One\_Sided\_DLS}} and \mbox{\texttt{Two\_Sided\_DLS}}.

The performance degradation of the DLS techniques with \mbox{\texttt{Two\_Sided\_DLS}} is due to mapping the master to a KNL core, which has the lowest processing capabilities~(cf. Section~\ref{sec:DoE}).
Recall that in \mbox{\texttt{Two\_Sided\_DLS}}, the master is responsible for serving work requests, and therefore, it has to divide the time between serving the work requests and performing its own chunks.
Therefore, if the master has a lower processing capabilities than the other processes, it becomes a performance bottleneck. 
Also, recall that \mbox{\texttt{One\_Sided\_DLS}} is designed to addresses this scenario~(Sections~\ref{sec:intro} and~\ref{sec:main}).
The coordinator process executes its own chunks, and is not responsible for the calculation and the allocation of the chunks to the other processes.

Figure~\ref{fig:1:2} shows that the DLS techniques with \mbox{\texttt{One\_Sided\_DLS}} perform comparably to their versions with \mbox{\texttt{Two\_Sided\_DLS}}.
For instance using the ratio 2:1, the \texttt{One\_Sided\_DLS} implementation of SS, GSS, TSS, FAC2, and WF required 108, 177, 125, 125, and 110 seconds, respectively. 
The \texttt{Two\_Sided\_DLS} implementation of the same techniques required 105, 175, 135.6, 125, and 106.45 seconds, respectively. 
Also, using the ratio 2:1, the DLS techniques behaved similarly regardless of their implementation approach.

\begin{figure*}[!t]
	\captionsetup[subfigure]{justification=centering}
	\begin{subfigure}{\textwidth}
		\centering
		\includegraphics[clip,trim=2cm 1cm 2cm 0cm,scale=0.55]{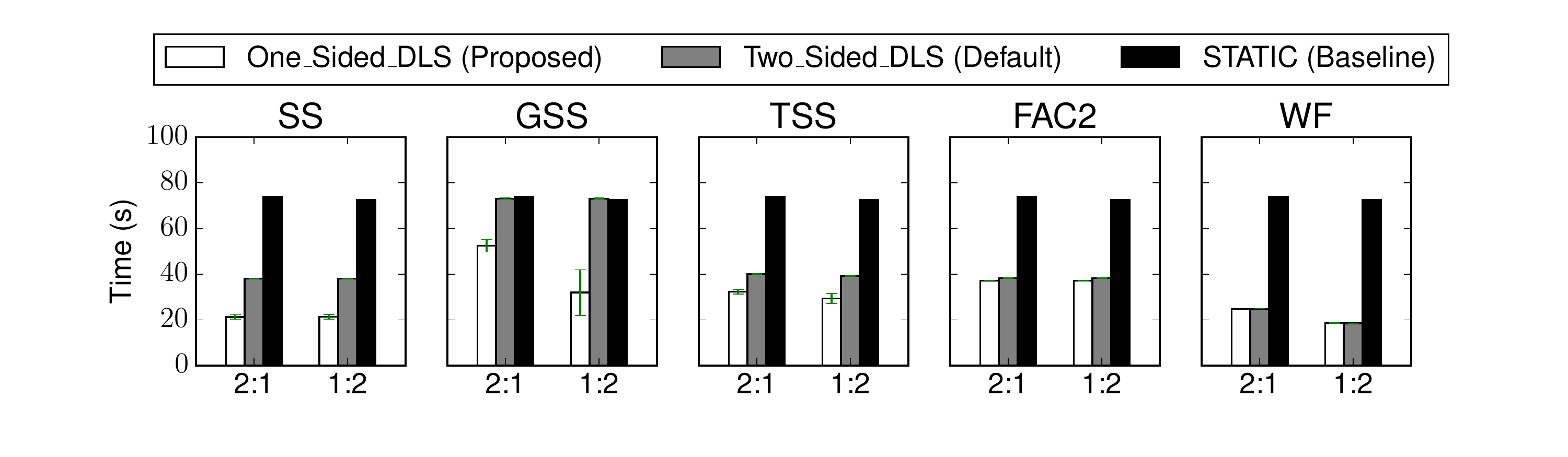}
		\caption{The coordinator$|$master is mapped to a KNL core}
		\label{fig:2:1}
	\end{subfigure}
	\\
	\begin{subfigure}{\textwidth}
		\centering
		\includegraphics[clip,trim=2cm 1cm 2cm 1cm, scale=0.55]{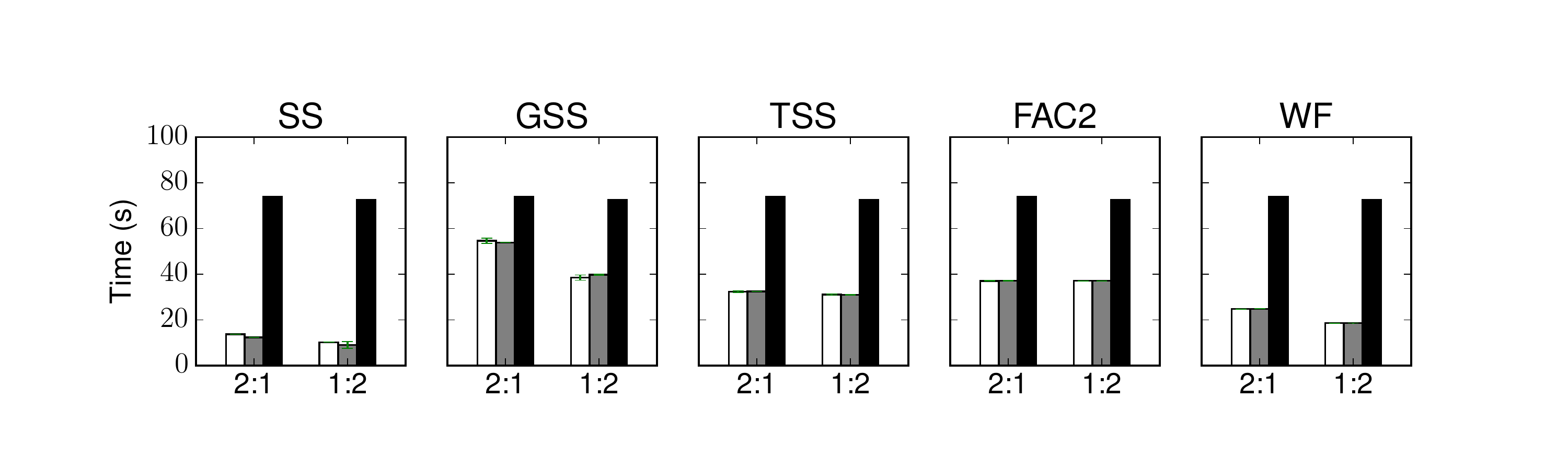}
		\caption{The coordinator$|$master is mapped to a Xeon core}
		\label{fig:2:2}
	\end{subfigure}
	\caption{Performance of the proposed approach vs. the existing \mbox{master-worker} based approach for the Mandelbrot set. The $x$-axis represents the two ratios between the KNL cores and the Xeon cores.}
	\label{fig:2}
\end{figure*}

For the Mandelbrot application, Figure~\ref{fig:2} confirms the same performance advantages of the proposed approach as for the PSIA application.
The DLS techniques implemented with \texttt{One\_Sided\_DLS} performed equally whether the coordinator was mapped to a KNL core or to a Xeon core.
The performance of certain DLS techniques with \texttt{Two\_Sided\_DLS} degraded when the master was mapped to a KNL core  compared to their performance when the master was mapped to a Xeon core.

\clearpage
\clearpage

Overall, Figures~\ref{fig:1} and \ref{fig:2} highlight two important observations.
 \emph{First observation:} The performance variation for executing a certain experiment using the \mbox{\texttt{One\_Sided\_DLS}} approach is higher than the performance variation when executing the same experiment using the \mbox{\texttt{Two\_Sided\_DLS}} approach.
The reason behind such variation is the manner in which concurrent messages are implemented at the MPI layer in \mbox{\texttt{One\_Sided\_DLS}} and \mbox{\texttt{Two\_Sided\_DLS}}.
In the current work, the Intel MPI is used to implement both approaches, \mbox{\texttt{One\_Sided\_DLS}} and \mbox{\texttt{Two\_Sided\_DLS}}.
Intel MPI uses the \textit{Lock Polling} strategy to implement \texttt{MPI\_Win\_lock} in which the origin process repeatedly issues \mbox{lock-attempt} messages to the \texttt{coordinator} process until the lock is granted~\cite{zhao2016scalability}.		
On the contrary, \mbox{\texttt{Two\_Sided\_DLS}} uses the \texttt{MPI\_Send}, \texttt{MPI\_Recv} and \texttt{MPI\_Iprobe} functions. 
For Intel MPI, in the case of simultaneous sends of multiple work requests to the master process, the master checks the outstanding work requests using \texttt{MPI\_Iprobe}, and serves them by giving priority to the request of the process with the smallest MPI rank.
The \mbox{\texttt{One\_Sided\_DLS}} has a high probability to grant the lock to different MPI processes at each trial, whereas, \mbox{\texttt{Two\_Sided\_DLS}} always prioritizes requests from the process with the smallest MPI rank.    
%This explanation justifies the high variation in the results obtained when GSS is implemented using \texttt{One\_Sided\_DLS} .
The GSS has the largest non-linear decrement among the decrements of the selected DLS techniques.
Therefore, GSS is \mbox{highly-sensitive} to the \mbox{chunk-assignment}.

\clearpage
\clearpage

\emph{Second observation}: FAC and WF exhibit a reduced sensitivity to mapping the master to a KNL or to a Xeon core.
This low sensitivity could be due to the \mbox{factoring-based} nature of these techniques.
Among all the assessed DLS techniques, FAC2 and WF assign chunks in batches, which increases the possibility for the master to have chunks of the same size as the other processing elements.
However, further analysis is needed to better understand such reduced sensitivity.

% !TEX root =  full_paper.tex
\section{Conclusion and Future Work}
\label{sec:conc}
A number of DLS techniques has been revisited and \mbox{re-evaluated} in light of and to enable them to benefit from the significant advancements in modern HPC systems, both at hardware and software levels. 
A distributed \mbox{chunk-calculation} approach~(\mbox{\texttt{One\_Sided\_DLS}}) has been proposed herein and is implemented using the MPI RMA and atomic \mbox{read-modify-write} operations with \mbox{passive-target} synchronization mode.
The \mbox{\texttt{One\_Sided\_DLS}} approach performs competitively against existing approaches, such as~\mbox{\texttt{Two\_Sided\_DLS}} that uses MPI \mbox{two-sided} communication and employs the conventional \mbox{master-worker} execution model.
\mbox{\texttt{One\_Sided\_DLS}} has the potential to alleviate the master-worker level contention of \mbox{\texttt{Two\_Sided\_DLS}} in \mbox{large-scale} HPC systems.
%The performance of existing implementations of DLS techniques~\mbox{\cite{LBtool,DLBLtool,Eleliemy:hpcc}} depends on the optimal choice of the processing element to execute the process that plays the role of the master.
%With the new proposed approach, the performance of the DLS techniques is negligibly unaffected by the arbitrary mapping of the \texttt{DATAOWNER} to any processing element in the system.

The present work revealed interesting aspects, planned as future work.
The performance of the two approaches considered herein, \mbox{\texttt{One\_Sided\_DLS}} and \mbox{\texttt{Two\_Sided\_DLS}}, is planned to be assessed with  additional applications. 
Specifically, the applications that require the return of the intermediate results upon the execution of each chunk of work.
These applications will help to assess the impact of the data distribution on the \mbox{\texttt{One\_Sided\_DLS}} approach.     
%In such cases, the performance of both, \mbox{\texttt{One\_Sided\_DLS}} and \mbox{\texttt{Two\_Sided\_DLS}}, is expected to degrade. 
%Yet, it remains of high interest to quantify this performance degradation.
%The performance assessment of the proposed approach with other MPI libraries is also of high interest for future work.
%Specifically, foMPI where other implementation strategies, e.g., \textit{Lock Queuing}, are employed to allow more scalable and memory efficient \mbox{one-sided} communications~\cite{foMPI}. 
%Therefore, examining the performance of the \texttt{DLS\_One} is of high interest due to the fact that it may be less sensitive the overhead associated with intermediate results compared to \texttt{DLS\_Two\_S} and \texttt{DLS\_Two\_M}.
%\flo{THIS SENTENCE ABOVE IS TOO SPECIFIC}
The scalability aspect of the proposed \mbox{\texttt{One\_Sided\_DLS}} approach also requires further study and analysis.

% !TEX root =  full_paper.tex
%\section{Reproducibility of This Work}
%\label{sec:repro}
%To ensure reproducibility of this work, apart from the information in Section~\ref{sec:DoE} about the application and the platform considered in this work, the code of the five DLS techniques using the two approaches is developed under the LGPL license, and is available online\footnote{https://drive.switch.ch/index.php/s/W2oYeCi9Z9Hsnvk}.
%The code was compiled with the Intel MPI compiler version 2018 update 1 with -O3 compiler optimization.
%The tag matching interface~(TMI) was selected as the low level implementation of the MPI library.
%All reported values represent the median of 20 repetitions of each experiment. 
%The job scheduling of the applications executing on the computing nodes used in this work is managed by Slurm version~17.02.7.
%Moreover, the raw results and the scripts used to generate all graphs that were presented in this paper are also available online with the code. 

\section*{Acknowledgment}
This work has been supported by the Swiss National Science Foundation in the context of the Multi-level Scheduling in Large Scale High Performance Computers” (MLS) grant number 169123 and by the Swiss Platform for Advanced Scientific Computing (PASC) project SPH-EXA: Optimizing Smooth Particle Hydrodynamics for Exascale Computing. 

%\balance
\newpage
\bibliographystyle{IEEEtran}
\bibliography{references}	
	
\end{document}